# Continuous Gaze Tracking with Implicit Saliency-Aware Calibration on Mobile Devices

Songzhou Yang, *Member, IEEE*, Meng Jin, *Member, IEEE*, and Yuan He, *Senior Member, IEEE*

**Abstract**—Gaze tracking is a useful human-to-computer interface, which plays an increasingly important role in a range of mobile applications. Gaze calibration is an indispensable component of gaze tracking, which transforms the eye coordinates to the screen coordinates. The existing approaches of gaze tracking either have limited accuracy or require the user's cooperation in calibration and in turn hurt the quality of experience. We in this paper propose vGaze, continuous gaze tracking with implicit saliency-aware calibration on mobile devices. The design of vGaze stems from our insight on the temporal and spatial dependent relation between the visual saliency and the user's gaze. vGaze is implemented as a light-weight software that identifies video frames with "useful" saliency information, sensing the user's head movement, performs opportunistic calibration using only those "useful" frames, and leverages historical information for accelerating saliency detection. We implement vGaze on a commercial mobile device and evaluate its performance in various scenarios. The results show that vGaze can work at real time with video playback applications. The average error of gaze tracking is 1.51cm (2.884°) which decreases to 0.99cm (1.891°) with historical information and 0.57cm (1.089°) with an indicator.

## 1 INTRODUCTION

Gaze reflects the potential intention and interest of a user on the content displayed on the mobile device. Gaze tracking is a useful human-to-computer interface, which plays an increasingly important role in a range of mobile applications, such as recommendation systems [1], [2], viewport-driven video streaming [3], [4], gaze-based human-to-computer interaction (HCI) [5], [6], etc.

The primary task of gaze tracking is to project the eye movements captured by the camera onto the screen of the mobile device. Due to the intrinsic mobility of the user, the relative position of the user's eyes to the screen may be varied now and then. Therefore, *gaze calibration*, which transforms the eye coordinates to the screen coordinates, becomes an indispensable component of gaze tracking.

Gaze calibration is a non-trivial task, as it concerns not only the tracking accuracy, but also affects the quality of experience of a mobile user. Early works mainly resort to *explicit calibration*, which requires the user's cooperation to gaze at stimuli points at predefined coordinates on the screen [7]–[12]. Calibration in this way inevitably interrupts the normal usage and hurts the user's experience. Deep learning based gaze tracking [13]–[16] doesn't require calibration. Instead, it needs to train a neural network model, which directly transforms the eye position in the captured image to the corresponding gaze position on the screen. The tracking accuracy of those approaches highly depends on the training process and is likely to degrade when generalized to different users and different contexts. Running intensive neural network models on the mobile devices is another obstacle.

Recent works propose to leverage visual saliency for *implicit gaze calibration* [17], [18]. The visual saliency is a kind of visual information (i.e., distinctive color, intensity, orientation, objects, etc.) that is contained in the frame itself. Compared to other content in the screen, visual saliency is much more likely to draw the user's attention. Hence, saliency is deemed as a significant indicator of user's gaze. More importantly, implicit gaze calibration doesn't interrupt the normal usage, so it can preserve high quality of experience during the whole process of gaze tracking. Unfortunately, directly or blindly using the saliency for calibration leads to poor gaze tracking accuracy.

The reason behind is the saliency is not always "usable" for gaze calibration. We look into the mechanism of human attention and find that the effectiveness of saliency is actually determined by both spatial and temporal features of the saliency. Specifically, the user's attention is in the *bottom-up mode* during the first around 150ms after a new frame comes into his/her sight. In that short period, the user's attention is driven to distinct regions in the frames. After that, the user's attention enters the *top-down mode*, where the user's consciousness dominates the gaze to semantic regions. Whether the saliency is suitable for calibration then depends on the saliency quality of the frame. For example, when a frame contains multiple distinct regions or contains relatively large distinct regions, the saliency inside generally has low effectiveness in drawing the user's attention. How to sufficiently and properly exploit saliency information for gaze calibration remains an open problem.

In order to address the above problem, we in this paper propose vGaze, continuous gaze tracking with implicit saliency-aware calibration on mobile devices. The design of vGaze stems from our insight on the temporal and spatial dependent relation between the saliency and the user's gaze/attention. vGaze is implemented as a light-weight

Corresponding author: Yuan He, E-mail address: heyuan@tsinghua.edu.cn.
E-mails of other authors: yangsz18@mails.tsinghua.edu.cn, jinm@sjtu.edu.cn

software that identifies frames with saliency information and filter "useful" saliency, senses the user's eye movement, and performs opportunistic implicit calibration using only those "useful" saliency. Also vGaze can leverage historical information for VOD data to enhance saliency. vGaze realizes accurate and efficient gaze tracking with such a saliency-based calibration, without sacrificing the quality of experience.

Our contributions are summarized as follows:

- vGaze is the first work that leverage temporal and spatial saliency for implicit calibration and quantifies the effectiveness of the visual saliency contained in frames. It reveals fundamental temporal and spatial relationship between the saliency and the user's gaze/attention.
- The design of vGaze tackles several critical challenges in saliency-aware gaze calibration: i) By tracing the temporal and spatial features of frames, vGaze enables opportunistic calibration, which answers the key questions, i.e., when and how to utilize saliency for calibration. ii) vGaze has high applicability: it includes a gaze compensation method to correct the gaze distortion caused by the biases camera position, a head movement tracking module and a scene cut monitoring module to trigger re-calibration.
- We implement vGave on commercial mobile device and evaluate its performance with extensive experiments. vGaze can work at real time with video playback applications. The average error of gaze tracking is 1.51cm.

The rest of this paper is organized as follows. Section 2 discusses the related works. Section 3 presents the preliminaries of our work. We elaborate on the design of vGaze in Section 4, implement it and evaluate it in Section 5. We conclude this paper in Section 6.

## 2 RELATED WORK

Based on whether the calibration process is required, we classify the existing gaze tracking methods into two categories: calibration-free methods and calibration-based methods.

### 2.1 Calibration-free methods

This kind of methods mainly leverage deep learning methods to directly infer the gaze without calibration. Specifically, the basic idea is to extract features from images of the eyes and map them directly to points on the gaze plane based on a deep learning model. The model is usually trained based on large-scale datasets of eye images and the corresponding ground truth of gaze positions [13]. For example, GazeCapture [14] trains a convolutional neural network based on an eyes image dataset which are captured using front cameras of mobile devices. Mayberry et al. [19] design a neural network training with data collected from cameras on eyeglasses to predict the user's gaze with such an eyeglass. Park et al. [15] design a deep neural network to generate an intermediate pictorial representation of eye, which is further used for gaze tracking. A problem of these methods is that their accuracy largely rely on the scale of the training data. Thus they suffer poor performance once the training data are insufficient. More importantly, such deep learning based method is difficult to be execute on the resource-limited mobile devices.

### 2.2 Calibration-based methods

Traditional methods will first compute the gaze direction based on the anatomy eye model. Then the intersection of the gaze direction and the gaze plane (e.g., the screen) determines the gaze position. Here, a calibration process is required to set some parameters (e.g., the relative position between the user's eye and the screen). For example, Ohno and Mukawa [20] utilize two cameras to obtain the eye position and utilize an infrared camera to build the reflection model of the eyes. To further calibrate the model, the user is asked to look at two stimuli on the screen. Mora and Odobez [21] utilize Kinect to combine 2D image information and 3D depth information to build a 3D head model and use this model to infer gaze direction. An offline calibration is needed to fit the user's features to the 3D model. Similar methods include [22], [23]. This kind of calibration is also widely used by commercial gaze trackers like Tobii [24], with the support of hardwares. Some works replace the camera with other electronic components, like photodiodes. LiGaze [25] analyses reflected screen light by user's eye with a ring of photodiodes placed on VR headsets to infer the user's gaze direction. Li and Zhou [26] use near-infrared light deployed on the eyeglasses as a light source along with photodiodes to capture changes in the light reflected by the user's eyes to infer the user's gaze. In these two works, initial calibration is used to eliminate user diversity. In summary, those methods either need explicit calibration for gaze tracking initialization or require dedicated hardware. So they are not suitable for mobile devices.

With the development of computer vision, methods of using image information have been proposed. These methods usually build a regression model leveraging the facial/eye features and the ground-truth gaze positions. Using the regression model, one can output estimated gaze positions with given facial/eye features. In this kind of methods, calibration process is required to collect data for regression analysis. The most popular calibration method is the 9-points calibration [7], [8], where a $3 \times 3$ visual point matrix is displayed on the screen and the user is asked to gaze at each point in turn. Obviously, the 9-points calibration usually takes a long time, which harms user experience. To solve this problem, Pfeuffer et al. [9] propose a smooth pursuit-based calibration method. Taking advantage of the human eye's automatic tracking of moving targets, this approach can collect more valid data in a shorter period of time.

However, although being accelerated, such explicit calibration process inevitably interrupt the continuous experience of the user. To solve this problem, Sugano et al. [17] leverage saliency information to achieve implicit calibration. In this method, gaussian process regression is used to learn the mapping between the images of the eyes and the gaze points. Sugano and Bulling [18] bring saliency into egocentric video. They use saliency extracted from the outer camera to calibration eye position from the inner camera. However, existing methods just blindly involves all the saliency information in the regression process, resulting in high calibration error since not all the frames contain useful saliency information for calibration.

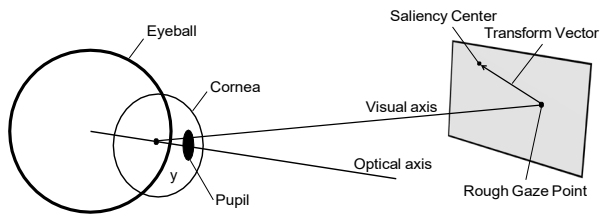

Fig. 1. Optical reflection model between the user's eye and the screen.

## 3 PRELIMINARY

In this section, we briefly introduce the principle of gaze calibration and gaze tracking, the basic knowledge of attention and saliency and our insight and process on saliency.

### 3.1 Principle of gaze calibration and tracking

The basic concept behind gaze tracking is to capture user's eye movement, and map it to the points on the gaze plane (i.e., the screen), as shown in Figure 1. Two important pieces of information are required in this process:

The first is a 3D model of the eye, based on which we can estimate the visual axis (gaze direction). Then, the intersection of the axis and the gaze plane determines the gaze point. Here, the 3D model of the eye can be captured by a RGB-D camera, which is widely available on today's smartphones such as iPhone X, Huawei Mate 20, OPPO Find X, etc. Specifically, an inferred projector projects a beam of structured infrared light onto the user's face especially the eye area and the RGB-D camera captures the reflection from the eye. The reflected structure light contains the depth information, based on which we can construct a 3D movement model of the eye. The advantage in using infrared light is that it is imperceptible to human eyes and is immune to the influences of environment light. Moreover, it protects the user's privacy.

The second is the relative position between the user's eye and the gaze plane. Without this information, the estimated gaze position suffers an offset from the real position. So, a gaze calibration process is required to compensate this offset. In typical calibration process, users are asked to fixate their gaze on certain stimuli on the screen, meanwhile the movement of their eyes is captured by the camera. The stimuli act as the ground truth of the gaze positions. The offset between the estimated gaze position and ground truth position captures a *transform vector*. Assuming that the relative position between the eye and the screen remains unchanged within a short period of time, we can directly applying the transform vector on the estimated gaze positions for gaze calibration.

However, as we have discussed previously, the explicit calibration process will harm the user experience, especially in the mobile scenarios where the re-calibration process should be triggered frequently to update the transform vector. To solve this problem, researchers try to use the information contained in the video frame itself (i.e., visual saliency) to perform gaze calibration in an implicit manner.

### 3.2 Saliency-based calibration

The basic insight behind saliency-based calibration is that the user's attention/gaze is usually attracted by several salient regions/objects on the screen when he/she is watching a video. Such salient regions/objects are collectively called as saliency, which stands out from its neighbors and can immediately draw the user's attention. Therefore, the positions of those salient regions/objects can be treated as the ground truth of user's gaze position, which helps to estimate the user's gaze. Today, with the development of computer vision, many effective methods have been proposed to detect the salient regions/objects in a video or a frame. Borji et al. [27] comprehensively reviews existing saliency detection mechanisms. These methods generally output a visual saliency heatmap, which shows the saliency of each pixel in a frame and can be regarded as the probability distribution of gaze. Readers can refer our previous work [28] for examples and more detailed information, where we've proved the opportunity to utilize the visual saliency for gaze calibration

### 3.3 Deep into saliency

We can not derive the user's gaze from the saliency in some scenarios. For instance, when the frame contains multiple salient regions/objects, the effect of using saliency for calibration is poor, because we cannot tell which region the user is looking at. On the contrary, the frame even may not contain saliency, e.g., an all black frame. Also, if the saliency is relatively large (e.g., a close-up frame), we are not able to determine which sub-region the user gazes at. In summary, it is not always possible to leverage the saliency in the spatial dimension to infer the user's gaze. We call this the spatial effectiveness of saliency.

Also, ignoring the temporal effectiveness of saliency is the other reason why the existing saliency-based calibration methods suffer. The visual saliency is temporally related to the user's attention. There are two pathways for human visual attention, the bottom-up and the top-down. In the bottom-up visual pathway, the information presented in the brain is the original physical characteristics of the external stimuli transmitted through the visual pathway, including color, intensity, orientation, etc. In summary, the bottom-up visual attention is driven by the external environmental information. As for the top-down pathway, it refers to the higher-level joint cortex of the brain, including the prefrontal cortex (PFC) and the posterior parietal cortex (PPC) to carry out information in the visual pathway based on the goals of the current task and past knowledge. This is attention driven by information inside the brain. Whether the two are identical depends on the visual content. We refer the interested reader to [29].

Specific to videos or frames watching, once a new scene appears, the user will first be driven by bottom-up attention to pay attention on the distinct region in the frame, and then will be dominated by top-down attention to focus on semantic objects based on past knowledge. We gives an intuitive user study in [28].

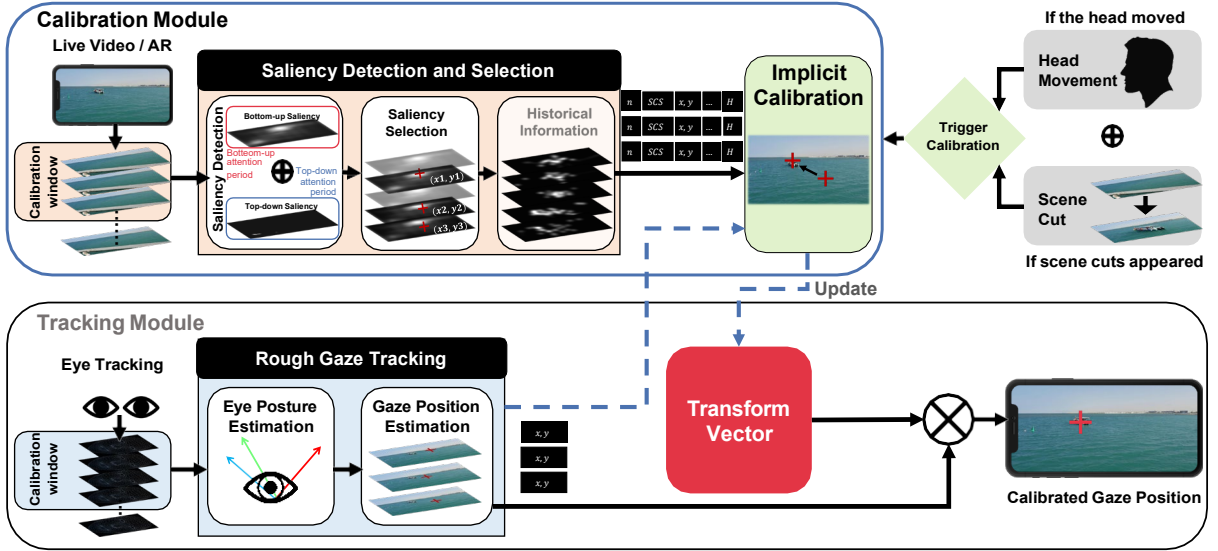

Fig. 2. Overview and Pipeline of the vGaze Design.

## 3.4 Spatiotemporal use of saliency

Knowing the effectiveness of saliency in spatial and temporal dimension, the next step is how to better utilize saliency based on this insight.

In order to solve aforementioned problem in spatial dimension, we design a metric to quantify the concentration degree of the saliency heatmaps, and only the saliency heatmaps with high concentration degree can be qualified as good calibration opportunities.

The concentration degree of a saliency heatmap is determined by two features as we mentioned above. One is the number of salient regions/objects on the heatmap. The other one is the area of salient region/object. Based on this, we propose a saliency metric called Saliency Concentration Score (SCS), which can be calculated as follow:

$$S = \begin{cases} \frac{1}{\log_2(n+1)} 1 - \frac{A_S}{A_T}, & n > 0 \\ 0, & n = 0 \end{cases} \quad (1)$$

where $n$ denotes the number of salient regions/objects in a frame. $A_S$ and $A_T$ are the areas of salient region/object and the whole frame, respectively. The value of SCS varies between 0 and 1. The less $n$ and the ratio of $A_S$ and $A_T$ are, the SCS value becomes closer to 1; and vice versa.

Our next target is to extract features $n$ and $A_S$ from each frame. To do this, we first binarize the heatmap of each frame to filter out the background pixels whose saliency value is lower than a threshold. The threshold of binarization is 128 in our later implementation. The remaining pixels reflect the salient regions/objects. Clearly, the ratio of the remaining pixels gives the ratio $\frac{A_S}{A_T}$. The number of regions/objects $n$ can be calculated by performing the connected component analysis on the binarized heatmap. Using SCS score, we can spatially select saliency. Concretely, we leverage above-mentioned Apple's algorithm [30] for bottom-up saliency detection. As for top-down saliency, we choose $U^2$-Net [31] to detect salient object.

Temporally, the bottom-up attention and top-down attention are handed over at the 100 millisecond level, more specifically 150 ms [32]. During saliency detection in later design, to match the change of attention, we leverage bottom-up saliency within first 150 milliseconds (which is about the length of 5 frames in a 30FPS video), then turn to top-down saliency in order to better match the attention mechanism. As for the recognition of scene cut, we rely on detecting key frames. Because during video encoding, once a scene cut occurs, it will be encoded as a key frame. Therefore, the key frame covers all scene cuts. By comparing the key frame with the previous frame, we can detect if there is a scene cut. We use pHash [33] to hash frames and calculate distance for detection in our design. By this, we can temporally select appropriate kind of saliency to denote the user's visual attention.

## 4 SYSTEM DESIGN

### 4.1 Overview

vGaze is a gaze tracking method which achieves highly reliable and accurate gaze tracking even in mobile scenarios. The key in achieving this is our saliency-aware calibration technique, which can perform gaze calibration whenever it's needed in an implicit manner. This implicit calibration is based on our insight about the temporal and spatial properties of saliency.

Figure 2 shows the workflow of vGaze. vGaze tracks the user's eye movement using a RGB-D camera and roughly projects it on the screen coordinate. By this, we get the rough gaze position estimation. Then, we compensate the rough estimation with a transform vector to get calibrated gaze position, which is acquired in calibration process. On the other hand, the rough estimation is used as an input for calibration when the calibration is required. The calibration is opportunistic process, which can be called based on monitoring head movement and scene cut. We perform calibration once the user's head moves or a scene cut occurs, where the two kinds of opportunity are independent. During calibration process, saliency information is extracted for frames in the calibration window. Based on our knowledge of the temporal and spatial dimensions of saliency, we select saliency in these two aspects. Specifically, we leverage bottom-up saliency and top-down saliency to match changes of user attention in the temporal dimension. Then, we filter out low-quality saliency maps by measuring the spatial characteristics of saliency. After selection, appropriate frames are used for the saliency-aware calibration to generate transform vector by comparison with rough gaze estimation acquired in tracking process. Here, optional historical information can be used for VOD data.

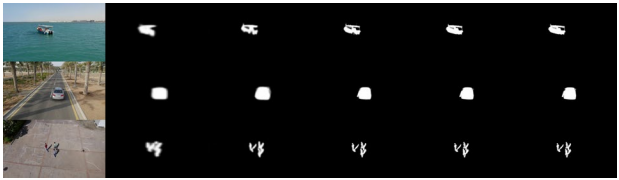

Fig. 3. Original frames and generated saliency maps of different frames: $68 \times 68$, $160 \times 90$, $320 \times 180$, $640 \times 360$ and $1280 \times 720$(Original), respectively. The shown saliency maps are stretched back to original resolution from original outputs.

The following of this section elaborates on the above components, providing the technical details of vGaze.

Here, we propose an implicit calibration mechanism in vGaze by leveraging the visual saliency. The inputs of calibration are frames from currently played video or AR content and the corresponding rough eye position estimations. vGaze use a calibration window to segment the frames and the eye position estimates for calibration. Each window involves $N$ frames $F_1$, $F_2$, ..., $F_N$ and $M$ eye position estimations $E_1$, $E_2$, ..., $E_M$. To eliminate the eye positioning errors, vGaze uses multiple eye positions in one frame. That is to say, the sampling rate of eye positions is higher than the frame rate (i.e., $M > N$) in vGaze.

### 4.2.1 Saliency Detection

The visual saliency heatmaps can be treated as probability distributions for gaze positions, which provides an opportunity for implicit gaze calibration. As we have discussed in Section 3.4, we leverage two kinds of saliency to denote the user's attention. Moreover, we dynamically select the appropriate detection algorithm according to different timings.

Before feeding each frame to the detection algorithms, we should first resize the frame into lower resolution ($68 \times 68$ in our implementation). The reason behind is two-fold. First, processing frames with high resolution e.g., 4K ($3840 \times 2160$) incurs high CPU, GPU, and energy overhead, which is unaffordable for resource-limited mobile devices. Second, since the resolution of the videos varies, we cannot predict the resolution of every video in advance. So, resizing all the frames to a fixed resolution is an effective and efficient solution to this problem. Note that reducing the resolution of the frame will not affect the saliency detection accuracy. This is because that the features of a frame that used in saliency detection (i.e., color, intensity, orientations, shape of objects, etc.) will not be changed at lower resolution.

The resized frames will then be fed to the visual saliency detection component to generate visual saliency heatmaps. Each heatmap is then normalized to a fixed range to maintain consistency.

To attest aforementioned process, we perform an experiment on random picked frames from videos we used in Section 5. We resize original frames (with $1280 \times 720$ resolution) into four different resolution: $640 \times 360$, $320 \times 180$, $160 \times 90$ and $68 \times 68$ respectively. Then each resolution is fed into saliency detection module to produce saliency map, Figure 3 demonstrates part of the results. The results show that there are merely tiny differences among different resolutions. Although the heatmap get a bit vague in low resolution, the saliency is still consistent with the high resolution which will not influence the result for calibration.

### 4.2.2 Saliency Selection

We've performed one selection to determine the type of saliency for each frame before saliency detection. After this temporal selection and saliency detection, there is still a problem, which is not all the frames can provide good opportunity for implicit calibration.

In Section 3.4, we propose a metric called SCS, which spatially quantify the effectiveness of saliency. Here, we use this metric to select frames that can be used for calibration. After calculating, we filter out the frames with low SCS value (0.6 in our implementation). We then extract saliency information from the remaining frames. Specifically, for each frame $F_i$, we find the pixel with highest saliency value in each connected component domain, whose coordinate denotes the position of the corresponding salient region/object. We then compress the coordinates of all the $n_i$ regions/objects on $F_i$, along with the region/object number $n_i$ and the SCS value $SCS_i$, into a feature vector $V_i$ as follows:

$$V_i = (n_i, SCS_i, x_1, y_1, ..., x_{n_i}, y_{n_i}) \quad (2)$$

The feature vector of each frame $F_i$ will be fed to the calibration component for implicit gaze error compensation.

Through both temporal and spatial selection, we now have suitable saliency maps that can be used for calibration.

We further perform an experiment to demonstrate the effectiveness of SCS in quantifying the calibration opportunity under each video frame. We use the six videos from EyeTrackUAV [34] dataset. The calibration window is set at 30, which means that we need to collect 30 valid frames to perform one calibration. In this experiment, we observe how many frames are required to collect enough valid frames (i.e., 30 valid frames) for calibration with different parameters, i.e., different binarization threshold (128 and 170) and different SCS threshold (0.6 and 0.8). The experimental results are shown as Figure 4.

The results show that with appropriate parameters, we are able to achieve high efficiency. We get the best performance when the threshold of binarization process is set at 128 and the SCS threshold is set at 0.6. In this case, the five videos averagely cost 34.46, 30.00, 30.43, 75.39, 43.47 frames to acquire 30 valid frames, respectively. The differences among videos are caused by the diversity in the concentration degree of the videos. For example, the video boat6 shows a boat sails on the sea, so the user's gaze will be largely concentrated on the boat with few interfering items. In this case, almost all the frames can be used for gaze calibration, so the average cost is 30.43 frames. This also prove that vGaze can successfully select

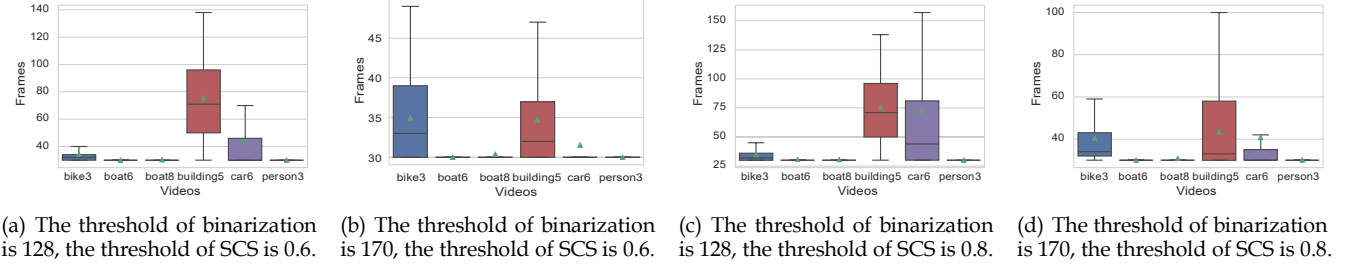

(a) The threshold of binarization is 128, the threshold of SCS is 0.6.

(b) The threshold of binarization is 170, the threshold of SCS is 0.6.

(c) The threshold of binarization is 128, the threshold of SCS is 0.8.

(d) The threshold of binarization is 170, the threshold of SCS is 0.8.

Fig. 4. Average frames cost for selecting valid 30 frames with different thresholds.

out the high quality frames. As for the video building6, the main contents in this video are different buildings. So the saliency is always distributed on those buildings. This is why the average cost is 75.39 frames, meaning that 50% of the frames are filtered out by vGaze. We further show how this design helps in improving the effective and efficient of the calibration process in Section 5.

### 4.2.3 Historical Information

Historical information about saliency can be used in some scenarios, for example, videos on demand (VOD). In this case, the input remains exactly the same among different users, so we can perform saliency detection and selection during the process of the first user's interaction or perform the calculation offline. The saliency information is recorded during this process. As a result, vGaze is able to reduce cost by replacing saliency detection and selection with recorded data, which reduces the cost of heavy online computing. Moreover, the historical user gaze trajectories act as an additional information which indicates the "statistical saliency", i.e., areas that most users focus on in practice. vGaze can further improve the accuracy of saliency by combining gaze trajectories and calculated saliency. Concretely, we denote user $p$'s the gaze point in the $k$th frame of the video as $T_{pk} = (x_{pk}, y_{pk})$, the statistical areas of interest for the $k$th frame are composed as

$$\mathbf{H}_k = \{T_{1k}, T_{2k}, ..., T_{nk}\} \tag{3}$$

where $n$ is the number of user trajectories. Now, we add this statistical information to the feature vectors.

$$V_i = (n_i, SCS_i, x_1, y_1, ..., x_{n_i}, y_{n_i}, \mathbf{H}_i) \tag{4}$$

Then the statistical information can be processed the same as saliency information in latter process.

### 4.2.4 Calibration

With extracted visual saliency vectors $(v_1, v_2, ..., v_N)$, we now can calibrate the error in rough gaze tracking result $\{G_1, G_2, ..., G_M\}$, which we will introduce in Section 4.3.1.

Before the calibration process, we first pre-process the rough gaze tracking result to filter out two sources of noise in the results.

The first source of noise is caused by the blink event. Specifically, the human eye generally blink 15-20 times in one minute [35]. The gaze position will rapidly change when the user blink his eyes. Similar phenomenon can be also observed in the saccade event. However, the gaze patterns are different in these two events. For the blink event, the gaze position rapidly changes but soon back to the original position. While, the gaze moves from one position to another position without moving back in the saccade. So, we can use z-score to eliminate the blink-trigged outliers while maintain the saccade event. Here, the z-score is calculated as:

$$z = \frac{x - \mu}{\sigma} \tag{5}$$

where the $x$ is the original value, $\mu$ is the mean of the whole values, and $\sigma$ is the standard deviation. We calculate the z-score for each rough gaze position in the calibration window and identify the rough gaze position as an outlier if the absolute value of its z-score is greater than a threshold $a$. We empirically set the threshold at 3 in our design.

Besides the blink event, error in eye positioning also brings noise in gaze tracking result. Specifically, it incurs small jitters in the rough gaze positions. To filter out such jitters, we sample more than one eye positions for each frame. We calculate the average of the $n_i$ rough gaze positions $\{G_1, G_2, ..., G_{ni}\}$ corresponding to one visual saliency vector $v_i$. The averaged gaze position is denoted as $\hat{G}_i$. Since we use $N$ frames in one calibration window, we will get $N$ average gaze positions $\{\hat{G}_1, \hat{G}_2, ..., \hat{G}_N\}$.

Now, we can use the $N$ gaze positions $\{\hat{G}_j\}$ and the $N$ saliency vectors $p$ for gaze calibration. Here, $N$ is set as 10 in implementation. Although one single saliency vector can be used to predict the real gaze positions, the accuracy is still not good enough to determine exact positions of gaze points because one single frame may contain multiple salient regions/objects. In addition, the uncertainty in user's behavior also incurs fluctuation for one single frame. For example, the user's attention can sometimes be attracted by unsalient area in the background. So we utilize the sequence of the $N$ frames to eliminate such error. Specifically, we first cluster $\{v_j\}$ and $\{\hat{G}_j\}$ separately. Figure 5 illustrate an example of the clustering result. Then for the clustering result of both $\{v_j\}$ and $\{\hat{G}_j\}$ (as shown by Figures 5a and 5b, respectively), we select the clusters with the most samples to denote the most frequent salient region and corresponding rough gaze region, respectively. By calculating the offset between the centroids of the two clusters, we get a vector called calibration transform vector $V_c$. Using this vector, we can compensate the error in the rough tracking result.

## 4.3 Continuous Tracking

### 4.3.1 Tracking

Once calibrated, vGaze is able to perform complete gaze tracking. Many existing works suffer serious privacy issues because they capture the user's face picture with the RGB camera of the mobile devices. To guarantee the user's privacy, we utilize the RGB-D camera in our design of vGaze. Thus, we only capture the depth and movement information instead of the user's face images. Specifically, we first capture the user's 3D eye information $E_1$, $E_2$, ..., $E_M$ using the front RGB-D camera, based on which we can get gaze direction. The intersection of the gaze direction and the gaze plane (the screen) determines the gaze points $\hat{G}_1$, $\hat{G}_2$, ..., $\hat{G}_M$ (which we also used in calibration process). However, a problem we face here is the position of the gaze plane (in relative to user's eyes) is not known. In the design of vGaze, we estimate the relative position using visual-inertial odometry (VIO) algorithm by by integrating the inertial information of the phone from inertial measurement unit (IMU) along with visual information from the camera's point of view.

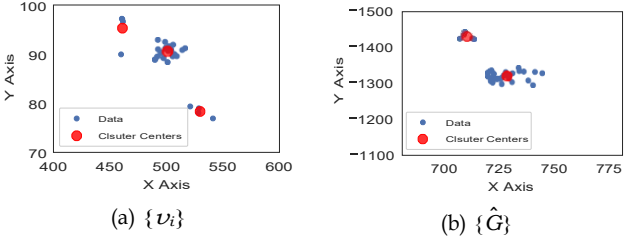

(a) $\{v_i\}$      (b) $\{\hat{G}\}$

Fig. 5. Clustering illustration of $\{v_i\}$ and $\{\hat{G}\}$ during calibration.

However, a challenge will arise when the user holds the phone with landscape posture. In this case, the camera will be rotated 90 degree to either the left side or the right of the user's face. As a result, the user's face is rotated a little when captured by the camera. The estimated gaze direction is also distorted, especially for the eye in the opposite direction of the camera. So, how to eliminate this distortion?

Without loss of generality, let's first consider the case where the camera is rotated to the user's left side. On the screen coordinate system, we assume that the origin locates at the bottom left corner of the screen. We have two key observations: i) the farther away the gaze is from the origin along x-axis, the larger its displacements will be when the eyeball rotates with a certain degree; ii) the nearer the gaze is from the origin along y-axis, the smaller its displacements will be when the eyeball rotates with a certain degree. The first observation is caused by the rotation of the user's face when captured by the camera. The second observation is caused by the oval structure of the eyes. With the oval structure the eyeball's rotation from left to right is more obvious than its rotation from upward to downward, even if the eyeball rotates with the same degree. Moreover, the eye opens wider when the user look upward compared with looking downward. So it is more difficult for the camera to capture the user's eye when he/she looks downward.

To compensate the distortion on $x$ axis, we compensate the $x$ value of the gaze position with the rotation of the user's face, which is captured by the front-facing camera of the mobile device. For the distortion on $y$ axis, we compensate the $y$ value of the gaze position with a constant value when $y$ is less than a threshold. With this compensation, we can perform gaze tracking in both portrait and landscape posture.

Finally, we compensate transform vector onto the rough gaze tracking result to acquire calibrated tracking result.

### 4.3.2 Recalibration

With the transform vector $V_c$, we can apply this vector on the the rough gaze position to acquire the calibrated gaze position as we just introduced. Now, a missing piece is when should we trigger such a calibration process. Our calibration mechanism is opportunistic, which is reflected in two independent aspects:

On the one hand, a transform vector works only if the user and the device hold a fixed relative position. Thus, we re-calibrate when a change in the relative position is detected. Otherwise, vGaze can directly use the previously calculated transform vector to perform calibrated tracking. We achieve such detection by tracking the user's face movement with the front-facing RGB-D camera. Specifically, when the relative position between the user and the device changes, the user's face posture captured by the camera will inevitably change. So, we continuously capture 3D information of the user's face during the gaze tracking process. Once the distance between two successively face postures is more significant than a pre-defined threshold (0.005 as default), a change in the relative position is detected. Then a new calibration process is triggered to update the transform vector. This recalibration could also happen during an existing calibration process to maintain the calibration quality. The visual content determines the type of saliency used in the re-calibration. The bottom-up saliency is used if the movement happens during the scene cut. Otherwise, the top-down saliency is used. In the former situation, the following re-calibration will be skipped.

On the other hand, we perform calibration when scene cuts appear. As we mentioned in Section 3.4, the bottom-up attention will immediately dominate the user's gaze after a scene cut. This kind of subconscious behavior has substantial confidence that connects the user's gaze with bottom-up saliency. Hence, we trigger calibration for detected scene cuts to maintain the calibration quality. In this kind of calibration, the length of the calibration window is defined as 5 frames to maintain consistency with the duration of attention.

## 5 IMPLEMENTATION & EVALUATION

In this section, we evaluate the performance of vGaze in various scenarios. We adopt 10 videos for evaluation. The videos are from EyeTrackUAV [34] dataset. Table 1 gives the details. Also, we invite volunteers to participate in our evaluation to assess the user diversity. There are 16 volunteers (10 males and 6 females) involved whose ages vary from 8 to 72 years old.

### 5.1 Implementation

In our implementation, we choose a iPhone Xs Max, which integrates Apple A12 Bionic of 2.49 GHz, 4GB RAM, 6.5-inches screen, TrueDepth camera and runs iOS 13.6 OS. The TrueDepth camera provides one kind of RGB-D cameras. Our implementation can apply to any iOS devices with TrueDepth camera like iPhone 11, iPad Pro and so on. Also our design of vGaze can be implemented on any Android

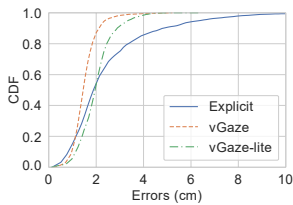
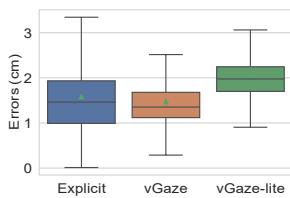
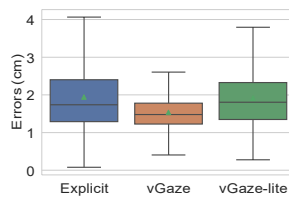
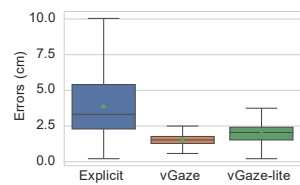

Fig. 6. Overall gaze tracking errors of vGaze and two baselines.

(a) Static      (b) Natural      (c) Dynamic

Fig. 7. Gaze tracking errors on 6 videos with 10 volunteers compared with explicit calibration and traditional saliency-based calibration in three different scenarios.

TABLE 1
Video details

| Title | Duration | Frames | Large Region | Multi Objects |
|-------|----------|--------|--------------|---------------|
| bike3 | 14s | 432 | | ✓ |
| boat6 | 27s | 804 | | |
| boat8 | 23s | 684 | | |
| building5 | 16s | 480 | ✓ | |
| car6 | 73s | 2194 | | |
| group2 | 89s | 2683 | | ✓ |
| person3 | 21s | 643 | | |
| person13 | 29s | 883 | | |
| truck4 | 42s | 1261 | ✓ | |
| wakeboard8 | 51s | 1543 | | ✓ |

* The Resolution and Sample Rates of all videos are 1280*720, 30FPS, respectively.

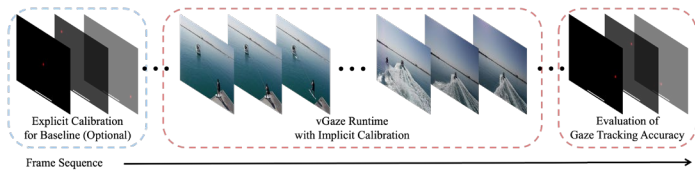

Fig. 8. Frame sequence of evaluation scenarios.

devices with RGB-D cameras, such as Huawei Mate 20, OPPO Find X, Honor Magic 2 and so on.

vGaze is coded with Swift and Objective-C++. In order to ensure the repeatability of frames between different users for evaluation, we use videos as the visual input in the implementation. This implementation can be easily converted to AR scenarios with simple settings. To acquire the RGB-D camera data we use ARKit framework [36]. OpenCV for iOS [37] is utilized for several frame processing functions.

## 5.2 Evaluation Scenarios

The implicit calibration approach in vGaze removes the existing of stimuli and the modification of the visual content to provide the user with continuously satisfying quality of experience. While stimuli act as ground truth in experiments. In order to evaluate the performance of gaze tracking and compare with explicit calibration, we constructed evaluation scenarios. We consider this evaluation pipeline, 1) we first compose frames with explicit dots to perform explicit calibration as baseline, which is optional pipeline for baseline comparison 2) then we run vGaze with normal frames from videos to perform normal vGaze calibration, 3) after a while, we replace the normal frames with artificial frames with explicit stimuli to measure the gaze tracking performance of vGaze by asking volunteers to gaze at these stimuli. Figure 8 shows the evaluation pipeline. Our experiments have been approved by the Institutional Review Board (IRB).

## 5.3 Gaze Tracking Performance

In this experiment, we evaluate the performance of vGaze on gaze tracking accuracy. Here, we take explicit calibration and traditional saliency-based calibration as baselines. To simulate existing work, we simplify our saliency-related process while retaining our recalibration mechanism and note this as vGaze-lite. Existing work doesn't have any recalibration mechanism. We eviscerate the historical module and note it as vGaze for most evaluations for universality. The historical module is evaluated later in Section 5.4. The aforementioned evaluation scenarios with all three steps work here. Specifically, we perform an explicit calibration with 5 red dots located on the center and four corners on the screen before vGaze starts. The user is asked to gaze at 5 dots in given order, which lasts for 10 seconds. Then the implicit calibration is performed for vGaze and vGaze-lite. After 300 frames normal running, the evaluation frames appear, all three types of tracking are evaluated. These frames are the same as those used for explicit calibration. Performance of all three methods are synchronously evaluated. To be fair, we perform evaluation in three scenarios. In the static scenario, the user is asked to stay static at the beginning to maintain the best performance for explicit calibration. In the dynamic scenario, the user is asked to perform at least one head movement in the middle of the videos to simulate possible movement on mobile scenarios. Also, the user's actions are not constrained to indicate in the natural scenario.

### 5.3.1 Overall Error

Figure 6 shows the overall results. 6 videos and 10 volunteers are involved here. We can see the overall error is 2.43cm, 1.51cm and 1.99cm for explicit calibration, vGaze and vGaze-lite, respectively. The corresponding angular error is 4.642°, 2.884° and 3.801°. To better understand the results, Figure 7 shows the results on three different scenarios. Figure 7a gives the comparison of three methods in static scenario. The average errors of explicit calibration, vGaze and vGaze-lite are 1.58cm (3.018°), 1.47cm (2.808°), 2.06cm (3.935°) respectively. The tracking error of vGaze is equivalent to explicit calibration, but vGaze doesn't need a long time waiting for calibration. Meanwhile the error of vGaze-lite increase by 30.38% and 40.14% compared with explicit calibration and vGaze. This shows our insight on saliency works effectively. In the Figure 7b, we see the errors reach 1.94cm (3.706°), 1.54cm (2.942°) and 1.92cm (3.668°), respectively in the natural scenario. The error of explicit calibration significantly increase by 22.78%, while the other two maintain the similar errors as the static scenario. The

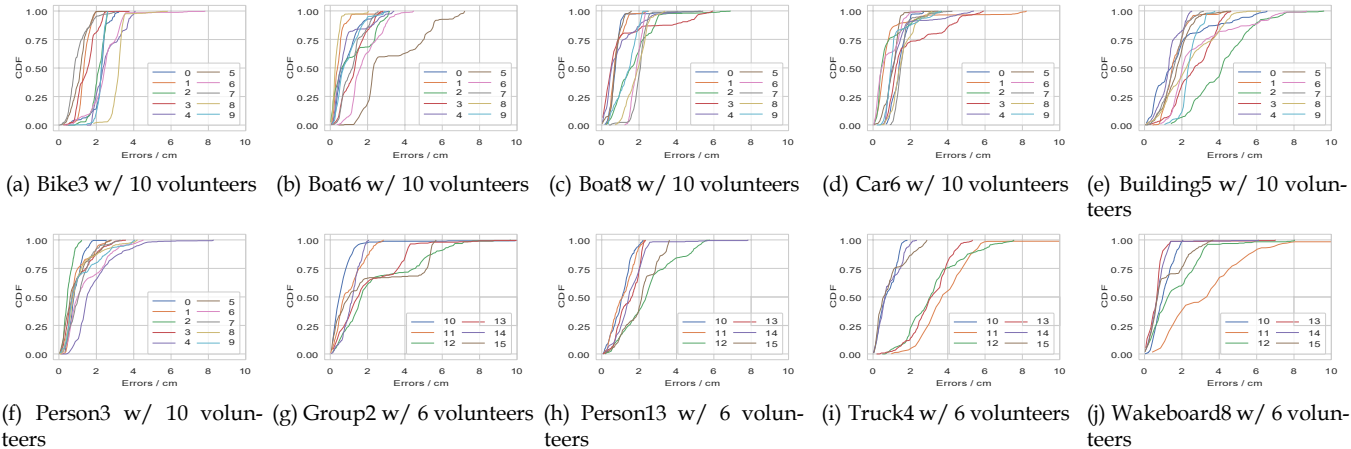

(a) Bike3 w/ 10 volunteers    (b) Boat6 w/ 10 volunteers    (c) Boat8 w/ 10 volunteers    (d) Car6 w/ 10 volunteers    (e) Building5 w/ 10 volunteers

(f) Person3 w/ 10 volunteers    (g) Group2 w/ 6 volunteers    (h) Person13 w/ 6 volunteers    (i) Truck4 w/ 6 volunteers    (j) Wakeboard8 w/ 6 volunteers reason behind is the tracking based on explicit calibration suffers once movement occurs. Also, this result proves the effectiveness of our recalibration mechanism. When it comes to the dynamic scenario, the explicit calibration perform worse. Figure 7c shows the result in this scenario. The average errors are 3.92cm (7.492°), 1.56cm (2.98°) and 2.00cm (3.82°), respectively. Since the user must move after calibration, the performance of explicit calibration is worse, increasing by 148.1% compared with the static scenario.

Fig. 9. Gaze tracking errors on different videos and volunteers

### 5.3.2 Errors of different videos

Further, Figure 9 represent more detailed results on different videos. The data come from two sets of experiments, one containing 6 videos and 10 volunteers (from our previous work [28]), and the other containing 4 videos and 6 volunteers. The average errors on 10 videos are 1.98cm (3.782°), 1.23cm (2.349°), 1.31cm (2.502°), 1.24cm (2.368°), 2.24cm (4.279°), 1.08cm (2.063°), 1.59cm (3.037°), 1.63cm (3.113°), 2.12cm (4.05°) and 1.42cm (2.712°), respectively. We see we reach best accuracy on Video Person3 with 1.08cm error and worst accuracy on Video Building5 with 2.24cm error. It's because of the differences on video contents. In Person3, the content is a person walking on a lawn where the user's attention can spontaneously focus on the walking person. As a result, the gaze tracking is accurate after calibration. It's the same for Boat6, Boat8, Car6 and Person13, where the gaze tracking errors are also relatively low. The content of Building5 is a sky view of some buildings where the saliency could be the whole buildings, which results in the calibration being relatively inaccurate even if vGaze has already filtered unsatisfying frames (Similar in Truck4, Bike3). As for Wakeboard8 and Group2, multiple moving objects exist. However, the accuracy doesn't distinctly fall due to vGaze's selection mechanism.

### 5.4 Historical Module

We evaluate the performance of historical module. We also leverage the gaze information from EyeTrackingUAV [34] to gather more historical information for eliminating random errors.

Figure 12 represents that the historical information reduce the errors. The overall errors decrease from 1.41cm to 0.99cm after introducing historical information. The historical information works especially well in complex frames like the building5 we mentioned before. The reason behind this is that historical information can compensate for the low quality of salience. In this example, the problem of too large saliency regions leading to difficult judgments is compensated by historical information.

### 5.5 Influence of Familiarity

Considering that the replaying of videos for the same volunteers will improve the users' familiarity of the videos. We re-evaluate the videos used in our previous work [28] with the same volunteers. We perform two continuous video playing for to the volunteers in this time. This experiment is two years after the last one, so the volunteers has low familiarity at the first playing and high familiarity at the second playing, while the videos are new to the volunteers two years ago. Figure 11 shows the result of this experiment. The errors were 1.47 cm, 1.42 cm and 1.37 cm when the familiarity was new, low and high, respectively. The errors are at the same level. The familiarity has almost no influence. The reason for this is two-fold. On the one hand, bottom-up attention is subconscious, which inevitably controls the user's behaviour. On the other hand, users may have "expectations" for the upcoming images after they are familiar with the video, leading them to still look at the areas.

### 5.6 Influence of the Mobile Phone Posture

The above experiments are carried out by placing the mobile phone on a phone stand in the landscape posture. In order to verify the effectiveness of our design on landscape posture and evaluate the performance of vGaze while the phone is held by the user. We conduct following experiments.

### 5.6.1 Landscape v.s. Portrait

We compare our design with two different scenarios, the portrait and landscape without any process. For the portrait scenario, videos are cropped to fit vertical screen without changing the contents. In these two scenario, they share the whole design of vGaze except the correction of distortion for landscape posture. We perform these two experiments in the static setting and compare them with above results of vGaze in static scenario.

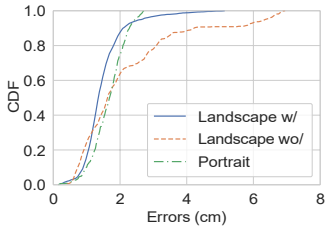

(a) Different phone orientation

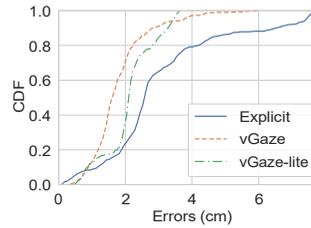

(b) Different phone placement

Fig. 10. Different phone postures.

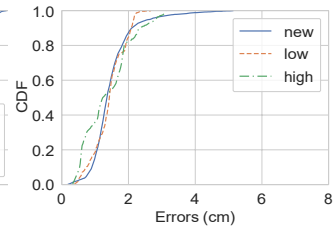

Fig. 11. Errors at different familiarity level.

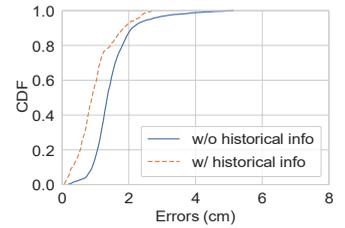

Fig. 12. Gaze tracking performance w/ historical information v.s. w/o.

Figure 10a gives the results. The average error of the portrait is 1.66cm, which is 12.93% higher than vGaze's 1.47cm error. The reason for this is that the oval structure of the eye makes the upward and downward rotation not obvious, which we discussed in Section 4.3.1 and solved for vGaze. As for the unprocessed landscape scenario, the error is 2.16cm increasing by 46.94% to vGaze's. The experimental results prove the effectiveness of our design for the landscape screen.

### 5.6.2 Fixed v.s. Holding

Here, we conduct experiments on two different use scenarios, the phone is placed on a stand and the phone is held by the user. We compare vGaze with explicit calibration and vGaze-lite. Figure 10b represents the result. The overall gaze tracking errors where the user holds the phone is 3.05cm, 1.79cm, 2.11cm for explicit calibration, vGaze and vGaze-lite, respectively. Compared to the errors 2.43cm, 1.51cm and 1.99cm where the phone is placed on a stand, the error increased in this scenario. Because this scenario is more dynamic, the relative position between the user and the screen has more opportunities to change. The explicit frames for collecting data have more chance to appear before recalibration and thus the recalibration is blocked, which results the gaze tracking inaccurate while data collecting. Also, the phone held by the user is not as stable as when the mobile phone is placed, which is more possible to generate errors while collecting data. However, vGaze will perform recalibration in time under the normal using scenarios.

### 5.6.3 Different Distance

We further compare our design in different using distance. In default evaluation, the distance between the user and the phone is 30cm. We expand the distance to 50cm and 80cm in this experiment. This experiment is conducted in the static setting and compared with the results of static scenario.

Figure 15 represents the results. The errors of distance 30cm and 50cm is 1.38cm and 1.37cm, respectively, which are even better than 1.47cm of default. The results show that vGaze is robust in different distance. The difference we think is because the system errors in different evaluations.

### 5.7 Influence of Capture Rate

In this experiment, we conduct experiment to explore if the capture rate of the camera influence the tracking accuracy. The default there the camera can averagely capture the eye movement twice every frame. We then set the camera to capture the eye movement every frame and every two frames. This experiment is conducted in the static scenario.

From Figure 16, there's almost no difference. The errors of the new settings are the 1.49cm and 1.44cm. The decrease of the capture rate means that there are less data used in a calibration window. The reason why the decrease does not influence the error might be the decrease of the amount of data doesn't influence the distribution of the data.

### 5.8 Influence of Parameters

There are several parameters in the design of vGaze. In order to evaluate them, we conducted controlled experiments. We use collected trajectories to perform offline calculation with different parameters in order to maintain consistency in different settings. The trajectories used is collected from dynamic scenarios. Four parameters are involved which are the threshold of binarization in SCS, the threshold of SCS, the frames used for calibration and the threshold of recalibration. Figure 13 represents the influence of different parameters.

### 5.8.1 Threshold of binarization

The threshold of binarization in saliency selection component influence the SCS value of each frame which is ultimately reflected in the calibration process. This threshold influences both the number of saliency and the area of saliency in the calculation of SCS. In general, low threshold doesn't filter the salient regions. As high threshold reduces the calculated area, but may increase the calculated number in complicated videos. Thus only middle values are selected for evaluation. The result shows the errors increase as the threshold rises.

### 5.8.2 Threshold of SCS

As for the threshold of SCS, low threshold results the less waiting time in calibration but introduces more unusable significance maps, so we only consider larger SCS values. That's the reason why the error increase when the threshold is reduced from 0.6 to 0.4. When the threshold is increased from 0.6 to 0.8, the error remains constant. It's because the SCS scores of selected frames under threshold with 0.6 are nearly all greater than 0.8.

### 5.8.3 Length of Calibration Window

The length of calibration window means how many frames are used for calibration. A short window will result in the presence of large randomness errors, while the user may move in a long calibration window. Hence, we choose moderate values for evaluation. The error increases when the threshold is increased from 10 frames to 30 because of more

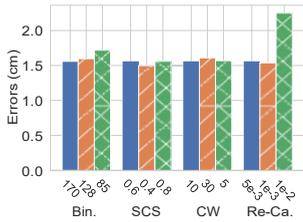

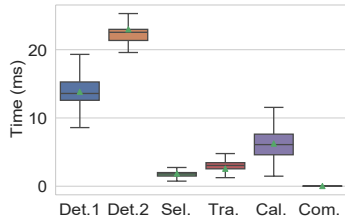

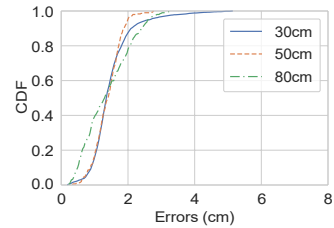

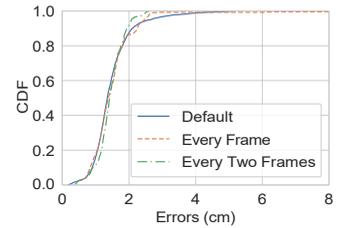

Fig. 13. Gaze tracking errors of vGaze with different parameters.

Fig. 14. Time elapsed by each module in vGaze.

Fig. 15. Different distance between the phone and the user.

Fig. 16. Different capture rate of camera.

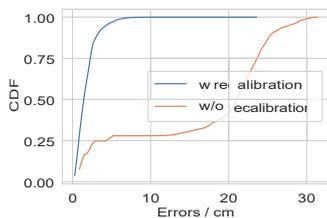

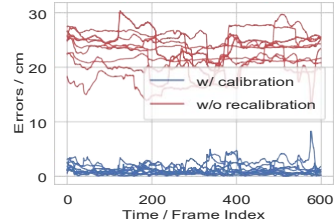

(a) Overall

(b) Errors along time

Fig. 17. Errors with and without calibration

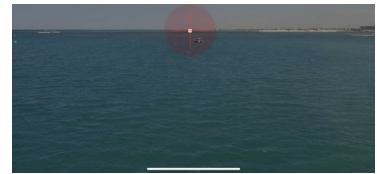

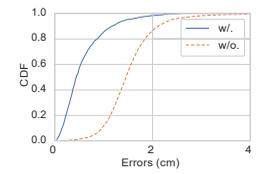

(a) Demonstration

(b) Exp. result

Fig. 18. Demonstration for interaction usage with an indicator and corresponding errors.

waiting time are needed. Also, the error increases when the calibration window is shortened. Maybe it's because the saliency of the frames used for calibration has minor changes where less frames cannot tolerate the changes.

### 5.8.4 Threshold of Movement Detection

When relative position between the user and the screen changes, a recalibration is needed to recover accurate gaze tracking. Radical strategy results more recalibration happening, but unnecessary recalibration may waste computing resource. Conservative strategy results less recalibration, but the accuracy of gaze tracking may suffer. The error get minor decrease with lower threshold (0.001), but the decrease is not significant. Because an appropriate threshold (0.005) has already handles most movement. However, the error get severe increase with higher threshold (0.010). Because the several continuous minor changes under threshold will compose a significant change which severely decrease the accuracy of gaze tracking.

Moreover, we perform an experiment to measure how often the relative position will change and what the error if there is no re-calibration. According to experiment the relative position changes 6, 18 and 159 times per minute in static, natural and dynamic scenario, respectively, when the threshold is set as 0.005. Figure 17b gives the result. The average error increases to 13.46cm when there is no recalibration mechanism. Figure 17 demonstrates the errors along time of video Person3 in evaluation phase. The results prove that the design of vGaze is effective and recalibration is necessary.

### 5.9 vGaze Efficiency

During experiments above, we simultaneously record the time elapsed by different components in vGaze. In Figure 14, we show times elapsed by fives components relevant to the frame process and gaze tracking. The abbreviations in the figure represent Saliency Detection (Bottom-up), Saliency Detection (Top-down), Saliency Selection, Rough Gaze Tracking, Calibration and Compensation of rough gaze tracking and transform vector. The average values are 22.98ms, 13.82ms, 1.83ms, 2.56ms, 6.25ms and 0.018ms respectively. The average detection time is 17.83ms. The average total time consumed by saliency detection and selection is 19.66 ms for a frame, which is much shorter than the frame display interval 33.33ms of 30 FPS video/ AR. According to the results, the maximum FPS supported by vGaze is about 50 FPS.

### 5.10 Interaction Usage

In the gaze-based interaction, there is usually an indicator on the screen to indicate the estimated user gaze, just like a cursor while using a mouse. This indicator, in turn, can also act as a potential stimulus, which provides additional information to improve the accuracy of gaze tracking.

Figure 18a gives the demonstration on using an indicator for interaction. With this indicator, the user is able to gaze more accurately on the target he/she want to interact with. We evaluate the gaze tracking error in this scenario, the result shows that the error decreases to 0.57cm which is only 37.75% of the error without indicator in Figure 18b. This result shows that the error of vGaze is totally acceptable for real-world usage.

## 6 CONCLUSION

In this paper, we explore how to achieve reliable gaze tracking on mobile device. With the insight of the temporal and spatial relation between the user's attention and the visual saliency, we present the design and implementation of vGaze, continuous gaze tracking with implicit saliency-aware calibration on mobile devices. vGaze opportunistically utilize the visual saliency information that naturally contained in frames to perform implicit calibration. We implement vGaze and evaluate its performance under different scenarios. The evaluation results show that vGaze realizes continuous gaze tracking with average 1.51cm (1.891°) errors without compromising the quality of user experience.

## ACKNOWLEDGEMENTS

This work is supported by the Smart Xingfu Lindai Project. We thank all the anonymous reviewers for their valuable comments and helpful suggestions.

## REFERENCES

[1] N. Silva, T. Schreck, E. Veas, V. Sabol, E. Eggeling, and D. W. Fellner, "Leveraging eye-gaze and time-series features to predict user interests and build a recommendation model for visual analysis," in *Proceedings of the ACM ETRA*, 2018.

[2] Q. Zhao, S. Chang, F. M. Harper, and J. A. Konstan, "Gaze prediction for recommender systems," in *Proceedings of the ACM RecSys*, 2016.

[3] Y. Guan, C. Zheng, X. Zhang, Z. Guo, and J. Jiang, "Pano: Optimizing 360 video streaming with a better understanding of quality perception," in *Proceedings of the ACM SIGCOMM*, 2019.

[4] S. Yang, Y. He, and X. Zheng, "Fovr: Attention-based vr streaming through bandwidth-limited wireless networks," in *Proceedings of the IEEE SECON*, 2019.

[5] T. E. Hutchinson, K. P. White, W. N. Martin, K. C. Reichert, and L. A. Frey, "Human-computer interaction using eye-gaze input," *IEEE Trans. on Systems, Man and Cybernetics*, vol. 19, no. 6, pp. 1527–1534, 1989.

[6] K.-N. Kim and R. Ramakrishna, "Vision-based eye-gaze tracking for human computer interface," in *Proceedings of the IEEE SMC*, 1999.

[7] C. Colombo and A. Del Bimbo, "Interacting through eyes," *Robotics and Autonomous Systems*, vol. 19, no. 3-4, pp. 359–368, 1997.

[8] C. H. Morimoto, D. Koons, A. Amit, M. Flickner, and S. Zhai, "Keeping an eye for hci," in *Proceedings of the IEEE SIBGRAPI*, 1999.

[9] K. Pfeuffer, M. Vidal, J. Turner, A. Bulling, and H. Gellersen, "Pursuit calibration: Making gaze calibration less tedious and more flexible," in *Proceedings of the ACM UIST*, 2013.

[10] F. Alnajar, T. Gevers, R. Valenti, and S. Ghebreab, "Calibration-free gaze estimation using human gaze patterns," in *Proceedings of the IEEE ICCV*, 2013.

[11] J. H. Goldberg and A. M. Wichansky, "Eye tracking in usability evaluation: A practitioner's guide," in *the Mind's Eye*, 2003, pp. 493–516.

[12] S. Baluja and D. Pomerleau, "Non-intrusive gaze tracking using artificial neural networks," in *Proceedings of the NIPS*, 1994.

[13] R. Stiefelhagen, J. Yang, and A. Waibel, "Tracking eyes and monitoring eye gaze," in *Proceedings of the PUis*, 1997.

[14] K. Krafka, A. Khosla, P. Kellnhofer, H. Kannan, S. Bhandarkar, W. Matusik, and A. Torralba, "Eye tracking for everyone," in *Proceedings of the IEEE CVPR*, 2016.

[15] S. Park, A. Spurr, and O. Hilliges, "Deep pictorial gaze estimation," in *Proceedings of the ECCV*, 2018.

[16] X. Zhang, Y. Sugano, M. Fritz, and A. Bulling, "It's written all over your face: Full-face appearance-based gaze estimation," in *Proceedings of the IEEE CVPR*, 2017.

[17] Y. Sugano, Y. Matsushita, and Y. Sato, "Appearance-based gaze estimation using visual saliency," *IEEE Trans. on Pattern Analysis and Machine Intelligence*, vol. 35, no. 2, pp. 329–341, 2012.

[18] Y. Sugano and A. Bulling, "Self-calibrating head-mounted eye trackers using egocentric visual saliency," in *Proceedings of the ACM UIST*, 2015.

[19] A. Mayberry, P. Hu, B. Marlin, C. Salthouse, and D. Ganesan, "ishadow: design of a wearable, real-time mobile gaze tracker," in *Proceedings of ACM MobiSys*, 2014.

[20] T. Ohno and N. Mukawa, "A free-head, simple calibration, gaze tracking system that enables gaze-based interaction," in *Proceedings of the ACM ETRA*, 2004.

[21] K. A. F. Mora and J.-M. Odobez, "Gaze estimation from multimodal kinect data," in *Proceedings of the IEEE CVPR*, 2012.

[22] H. Yamazoe, A. Utsumi, T. Yonezawa, and S. Abe, "Remote gaze estimation with a single camera based on facial-feature tracking without special calibration actions," in *Proceedings of the ACM ETRA*, 2008.

[23] L. Sun, Z. Liu, and M.-T. Sun, "Real time gaze estimation with a consumer depth camera," *Information Sciences*, vol. 320, pp. 346–360, 2015.

[24] "Tobii," https://www.tobii.com/.

[25] T. Li, Q. Liu, and X. Zhou, "Ultra-low power gaze tracking for virtual reality," in *Proceedings of the ACM SenSys*, 2017.

[26] T. Li and X. Zhou, "Battery-free eye tracker on glasses," in *Proceedings of the ACM MobiCom*, 2018.

[27] A. Borji, M.-M. Cheng, Q. Hou, H. Jiang, and J. Li, "Salient object detection: A survey," *Computational visual media*, pp. 1–34, 2019.

[28] S. Yang, Y. He, and M. Jin, "vgaze: Implicit saliency-aware calibration for continuous gaze tracking on mobile devices," in *Proceedings of the IEEE INFOCOM*, 2021.

[29] F. Katsuki and C. Constantinidis, "Bottom-up and top-down attention: different processes and overlapping neural systems," *The Neuroscientist*, vol. 20, no. 5, pp. 509–521, 2014.

[30] "Apple developer," https://developer.apple.com/documentation/vision/vngenerateattentionbasedsaliencyimagerequest.

[31] X. Qin, Z. Zhang, C. Huang, M. Dehghan, O. R. Zaiane, and M. Jagersand, "U2-net: Going deeper with nested u-structure for salient object detection," *Pattern Recognition*, vol. 106, p. 107404, 2020.

[32] C. E. Connor, H. E. Egeth, and S. Yantis, "Visual attention: bottom-up versus top-down," *Current biology*, vol. 14, no. 19, pp. R850–R852, 2004.

[33] C. Zauner, "Implementation and benchmarking of perceptual image hash functions," *Online*, 2010.

[34] V. Krassanakis, M. Perreira Da Silva, and V. Ricordel, "Monitoring human visual behavior during the observation of unmanned aerial vehicles (uavs) videos," *Drones*, vol. 2, no. 4, p. 36, 2018.

[35] A. R. Bentivoglio, S. B. Bressman, E. Cassetta, D. Carretta, P. Tonali, and A. Albanese, "Analysis of blink rate patterns in normal subjects," *Movement disorders*, vol. 12, no. 6, pp. 1028–1034, 1997.

[36] "Arkit," https://developer.apple.com/augmented-reality/.

[37] Opencv. [Online]. Available: https://opencv.org/

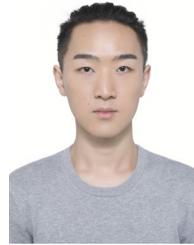

**Songzhou Yang** received the B.S. degree in School of Software from Tsinghua University in 2018, where he is currently pursuing the PhD's degree in software engineering. His research interests are in the field of extended reality (XR) and human-computer interaction (HCI).

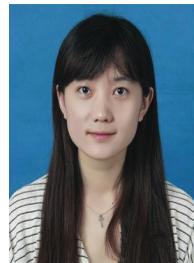

**Meng Jin** received the B.S., M.S., and Ph.D. degrees in computer science from Northwest University, Xi'an, China, in 2012, 2015, and 2018, respectively. She is currently a Post-Doctoral Researcher with the School of Software and BNRist, Tsinghua University. Her main research interests include backscatter communication, wireless network co-existence at 2.4 GHz, mobile sensing, and clock synchronization.

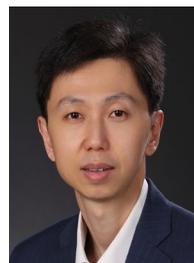

**Yuan He** is an associate professor in the School of Software and BNRist of Tsinghua University. He received his B.E. degree in the University of Science and Technology of China, his M.E. degree in the Institute of Software, Chinese Academy of Sciences, and his PhD degree in Hong Kong University of Science and Technology. His research interests include wireless networks, Internet of Things, pervasive and mobile computing. He is a member of the IEEE and ACM.